\input amstex.tex

\documentstyle{amsppt}
\def\version{BGV version 7c}

\topmatter
\author Thomas P. Branson${}^\dag$, Peter B. Gilkey${}^\ddag$, Dmitri V.
Vassilevich${}^*$\endauthor
\address \dag Mathematics Department, University of Iowa, Iowa City IA
    52242 USA. \endaddress\email: branson\@math.uiowa.edu\endemail
\address \ddag Mathematics Department, University of Oregon, Eugene Or
     97403 USA\endaddress\email gilkey\@math.uoregon.edu\endemail
\address *Department of Theoretical
     Physics, St. Petersburg University, 198904
     St. Petersburg Russia.\endaddress\email:
      vasilev\@snoopy.niif.spb.su\endemail
\title Vacuum expectation value asymptotics for
       second order differential operators on manifolds with
       boundary\endtitle
\thanks ${}^\dag$ Research partially supported by an
     international travel grant of the NSF (USA)\endthanks
\thanks ${}^\ddag$ Research partially supported by the NSF
     (USA) and MPIM (Germany)\endthanks
\thanks ${}^*$ Research partially supported by GRACENAS (Russia)
     and DAAD (Germany)\endthanks
\def\nmonth{\ifcase\month\ \or January\or
   February\or March\or April\or May\or June\or July\or August\or
   September\or October\or November\else December\fi}
\thanks \version\ printed\ \number \day\ \nmonth\ \number\year
   \endthanks
\def\pend{\operatorname{End}}\def\dvol{d\operatorname{vol}}
\def\ptr{\operatorname{tr}_V}\def\ord{\operatorname{ord}}
\def\PTR{\operatorname{Tr}_{L^2}}
\abstract Let $M$ be a compact Riemannian manifold with
    smooth boundary.
    We study the vacuum expectation value of an
    operator $Q$ by studying $\PTR Qe^{-tD}$, where $D$
    is an operator of Laplace type on
    $M$, and where $Q$
    is a second order operator with scalar leading
    symbol; we impose Dirichlet or modified Neumann
    boundary conditions.\endabstract
\endtopmatter
\rightheadtext{Vacuum expectation value asymptotics}
\leftheadtext{T. Branson, P. Gilkey, and D. Vassilevich}
\def\Arefb  {1.1}      

\def\BREFa  {2.1}      
\def\BREFc  {2.2}      
\def\BREFd  {2.3}      
\def\BREFg  {2.4}      
\def\BREFh  {2.5}      

\def\CREFa  {3.1}      
\def\CREFb  {3.2}      
\def\CREFc  {3.3}      

\def\DREFa  {4.1}      
\def\DREFb  {4.2}      
\def\DREFc  {4.3}      
\def\DREFd  {4.4}      
\def\DREFe  {4.5}      
\def\DREFg  {4.6}      
\def\DREFh  {4.7}      
\def\DREFi  {4.8}      
\def\DREFj  {4.9}      
\def\DREFk  {4.10}     

\def\EREFa  {5.1}      
\def\EREFb  {5.2}      
\def\EREFe  {5.3}      
\def\EREFg  {5.4}      
\def\EREFk  {5.5}      

\def\refBiDa   {1}     
\def\refBGa    {2}     
\def\refBGb    {3}     
\def\refBGc    {4}     
\def\refBGV    {5}     
\def\refFuj    {6}     
\def\refGia    {7}     
\def\refGib    {8}     
\def\refEOJ    {9}     

\def\Cur{F}\def\DF{{\Cal F}} 
\def\DR{{\Cal R}}            
\def\DL{{\Cal L}}            
\def\DN{{\Cal N}}            
\def\DB{{\Cal B}}            
\def\DD{{\Cal D}}            
\def\DE{{\Cal E}}            

\head\S1 Introduction\endhead

Let $M$ be a compact smooth Riemannian manifold of dimension $m$ with smooth
boundary
$\partial M$. We say that a second order operator $D$ on the space of smooth
sections
$C^\infty(V)$ of a smooth vector bundle over $M$ has {\it scalar leading
symbol}
if the leading symbol is $h^{ij}I_V\xi_i\xi_j$ for some symmetric 2-tensor $h$.
We
say that $D$ is of {\it Laplace type} if $h^{ij}$ is the metric tensor on the
cotangent bundle. Let
$D$ be an operator of Laplace type. If the
boundary of $M$ is non-empty, we impose  Dirichlet or Neumann boundary
conditions
$\DB $ to define the operator $D_\DB $, see \S4 for further details. Let $Q$
be an auxiliary second order partial differential operator on
$V$ with scalar  leading symbol; if the order of $Q$ is at most 1, then this
hypothesis is satisfied trivially. As
$t\downarrow0$, there is an asymptotic expansion
$$\PTR(Qe^{-tD_\DB })\sim\sum_{n=-2}^\infty
   a_{n}(Q,D,\DB )t^{(n-m)/2},\tag{\Arefb}$$
see Gilkey \cite{\refGib, Lemma 1.9.1} where a different
numbering convention was used. The invariants $a_n(Q,D,\DB )$ are locally
computable. We have $a_{-2}(Q,D,\DB )=0$ and $a_{-1}(Q,D,\DB )=0$ if $Q$
has order
at most 1. If the boundary of $M$ is empty, the boundary
condition $\DB $ plays no role and we drop it from the notation; in this case, if
$n$ is odd, then
$a_n(Q,D)=0$.

Our paper is motivated by several physical examples. First, consider a
Euclidean
quantum field theory with a propagator $D^{-1}$ depending on external fields.
Typically, $D$ is a second order differential operator of Laplace type. In the
one--loop approximation, the vacuum expectation value $<Q>$ of a second order
differential operator is given by $ <Q>=Tr_{L^2} (QD^{-1})$.
By formal manipulations, this can be represented in the form
$$
<Q> \sim \textstyle\int_0^\infty dt\ Tr_{L^2}(Qe^{-tD})\sim
\textstyle\int_0^\infty dt \sum_{n=-2}^\infty a_n(Q,D,{\Cal B})
t^{(n-m)/2}.
$$
These integrals are divergent at the lower limit and need to be
regularized. This can be done by replacing $0$ by $1/\Lambda$ in the
limits of integration; $\Lambda$ is called the ultraviolet cutoff
parameter. The coefficients $a_n(Q,D,{\Cal B})$ define the asymptotics
of $<Q>$ as $\Lambda\to\infty$. The first $m$ terms are divergent
and are essential for renormalization. The coefficients
$a_n(Q,D,{\Cal B})$ also define large mass asymptotics of $<Q>$
in the theory of a massive quantum field; for details see for example
\cite{\refBiDa}.

A second example is provided by quantum anomalies. In the Fujikawa approach
\cite{\refFuj}, the anomaly ${\Cal A}$ is defined as
${\Cal A}=\lim_{\Lambda\to\infty}Tr (Qe^{-D/\Lambda^2})$,
where $Q$ is the generator of an anomalous symmetry transformation, and
$D$ is a regulator. Usually divergent terms may be absorbed in
renormalization and one has that
${\Cal A}\sim a_{m-2}(Q,D,{\Cal B})$.
Other examples where these asymptotics arise naturally are the study of
the anomaly for an arbitrary local symmetry transformation, and in the
study of the vacuum expectation value of the stress-energy tensor.

In this paper, we will study the asymptotics $a_n(Q,D,{\Cal B})$ in a general
mathematical framework. In
\S2, we review the geometry of operators of Laplace type and derive some
variational
formulas. The operator $D$ determines the metric
$g$, a connection
$\nabla$ on $V$, and an endomorphism $E$ of $V$. Conversely, given these data,
 we can
define an operator of Laplace type $D(g,\nabla,E)$; see Lemma
\BREFd\ for details. Let $q_2$ be a symmetric 2-tensor and let $q_1$ be a
1-form
valued endomorphism of $V$. We shall use $q_2$ to define a variation of
the metric
$g(\varepsilon):=g+\varepsilon q_2$ and we shall use $q_1$ to define a
variation of
the connection $\nabla(\varepsilon):=\nabla+\varepsilon q_1$. Let
$Q_2:=\partial_\varepsilon D(g(\varepsilon),\nabla,E)|_{\varepsilon=0}$ and
let
$Q_1:=\partial_\varepsilon D(g,\nabla(\varepsilon),E)|_{\varepsilon=0}$. Let
$Q$ be a
second order operator with scalar leading symbol. We may decompose
$Q=Q_2+Q_1+Q_0$ for
$Q_0\in\pend(V)$ and for
$Q_2$ and $Q_1$ defined by suitably chosen $q_2$ and $q_1$.
Since $a_n(Q,D)=\sum_i
a_n(Q_i,D)$, it suffices to compute the $a_n(Q_i,D)$. In Lemma \BREFg,
we will study
the operators $Q_2$ and $Q_1$ and show that
$\partial_\varrho a_{n+2}(1,D(\varrho),\DB )=-
    a_n(\partial_\varrho D(\varrho),D(\varrho),\DB )$
for any 1-parameter family of operators of Laplace type and fixed
boundary condition $\DB $.
In \S3 and \S4 we use this variational formula and apply results of
\cite{\refBGa}
and \cite{\refBGV} to study the invariants $a_n(Q,D,\DB )$; in \S3 we consider
manifolds without boundary and in \S4 we consider manifolds with boundary.

An operator $A$ is said to be of Dirac type if $A^2$ is of Laplace type.
Branson and
Gilkey \cite{\refBGa} studied the asymptotics of $\PTR(Ae^{-tA^2})$ for an
operator $A$ of Dirac type on a closed manifold. In \S5, we use the
results of \S3 to
rederive these results and to compute some additional terms in the asymptotic
expansion. The numbering convention we shall use in this paper differs
from that used
in \cite{\refBGa}; the invariants $a_n(A,A^2)$ of this paper were denoted by
$a_{n-1}(A,A^2)$ in \cite{\refBGa}.

\head\S2 Geometry of operators of Laplace type\endhead

We adopt the following notational conventions. Greek
indices $\mu$, $\nu$, etc\. will range from $1$ through $m=\dim(M)$
and index local coordinate frames $\partial_\nu$ and $dx^\nu$ for
the tangent and cotangent bundles $TM$ and $T^*M$. Roman indices
$i$, $j$ will also range from $1$ through $m$ and index local
orthonormal frames $e_i$ and $e^i$ for $TM$ and $T^*M$. We shall suppress
the bundle indices for tensors arising from $V$. We adopt the Einstein
convention and sum over repeated indices. Let $D$ be an operator of
Laplace type. This means that we can decompose
$D$ locally in the form
$$D=-(g^{\nu\mu}I_V\partial_\nu\partial_\mu
   +a^\sigma\partial_\sigma+b)\tag{\BREFa}$$
where $a$ and $b$ are local sections of $TM\otimes\pend(V)$ and $\pend(V)$
respectively.
It is important to have a more invariant expression than
that which is given in equation (\BREFa). Let $\Gamma$ be the
Christoffel symbols of the Levi-Civita connection of the metric $g$ on
$M$, let $\nabla$ be an auxiliary connection on $V$, and let $E\in
C^\infty(\pend(V))$.  Define:
$$\eqalign{
&D(g,\nabla,E):=-(\operatorname{Tr}_g\nabla^2+E)\cr
=-g^{\mu\sigma}&\{I_V\partial_\mu\partial_\sigma
  +2\omega_\mu\partial_\sigma
  -\Gamma_{\mu\sigma}{}^\nu I_V\partial_\nu
  +\partial_\mu\omega_\sigma+\omega_\mu\omega_\sigma
   -\Gamma_{\mu\sigma}{}^\nu\omega_\nu\}-E.\cr}\tag{\BREFc}
$$
We compare equations (\BREFa) and (\BREFc) to prove the following Lemma.

\proclaim{\BREFd\ Lemma} If $D$ is an operator of Laplace type, then
   there exists a unique connection
   $\nabla$ on $V$ and a unique endomorphism $E$ of $V$
   so that $D=D(g,\nabla,E)$.\roster
\smallskip\item
   If $\omega$ is the connection 1-form of $\nabla$, then
   $\omega_\delta=g_{\nu\delta}(a^\nu+g^{\mu\sigma}\Gamma_{\mu
    \sigma}{}^{\nu }I_{V})/2$.
 \smallskip\item We have $E=b-g^{\nu \mu}(
    \partial_\mu\omega_\nu+\omega_\nu\omega
 _\mu-\omega_\sigma\Gamma_{\nu\mu}{}^\sigma)$.\endroster
\endproclaim

Let $D=D(g,\nabla,E)$. We  use the Levi-Civita
connection of the metric $g$ and the connection
$\nabla$ on $V$ to covariantly differentiate
tensors of all types. We shall let `;' denote multiple covariant
differentiation. Thus, for example,
$Df=-(f_{;kk}+Ef)$.

\proclaim{\BREFg\ Lemma} Let $D=D(g,\nabla,E)$ be an operator of Laplace
type. Let $q_2=q_{2,ij}$ be a symmetric 2-tensor and let $q_1=q_{1,i}$ be an
endomorphism valued 1-tensor.
Then\roster
\smallskip\item  $Q_1f:=
        \partial_\varepsilon D(g,\nabla+\varepsilon q_1,E)f|_{\varepsilon=0}
       =-2q_{1,i}f_{;i}-q_{1,i;i}f$.
\smallskip\item  $Q_2f:=\partial_\varepsilon D(g+\varepsilon q_2,\nabla,E)f
     |_{\varepsilon=0}
     =q_{2,ij}f_{;ij}+(2q_{2,ij;j}-q_{2,jj;i})f_{;i}/2$.
\smallskip\item Let $D(\varrho)$ be a smooth $1$-parameter family
of operators of Laplace type and fix $\DB $. Then
   $a_{n}(\partial_\varrho D(\varrho),D(\varrho),\DB )
   =-\partial_\varrho a_{n+2}(1,D(\varrho),\DB )$.
\endroster\endproclaim

\demo{Proof}
Fix a point $x_0\in M$; we may assume that $x_0$ is in the interior of $M$.
Choose
coordinates centered at $x_0$ and a local frame for
$V$ so that $g_{\nu\mu}(x_0)=\delta_{\nu\mu}$,
$\Gamma(x_0)=0$, and so that $\omega(x_0)=0$.
We use equation (\BREFc) to compute:
$$\eqalign{
&\partial_\varepsilon D(g,\nabla+\varepsilon q_1,E)(x_0)|_{\varepsilon=0}
     =-g^{\nu\sigma}(2q_{1,\nu}\partial_\sigma
     +\partial_\nu q_{1,\sigma})(x_0)\cr
&\qquad=(-2q_{1,i}\nabla_i-q_{1,i;i})(x_0),\cr
&\partial_\varepsilon \{g+\varepsilon
     q_2\}^{\nu\sigma}(x_0)|_{\varepsilon=0}=-q_{2,\nu\sigma}(x_0),\cr
&2\partial_\varepsilon\Gamma(g+\varepsilon
q_2)_{\nu\sigma}{}^\mu(x_0)|_{\varepsilon=0}
   =(q_{2,\mu\nu;\sigma}
   +q_{2,\mu\sigma;\nu}-q_{2,\nu\sigma ;\mu})(x_0),\cr
&\partial_\varepsilon D(g+\varepsilon q_2,\nabla,E)(x_0)|_{\varepsilon=0}\cr
&\qquad=(q_{2,\nu\sigma}(\partial_\nu\partial_\sigma
     +\partial_\nu\omega_\sigma)
     +\partial_\varepsilon
      \Gamma(g+\varepsilon q_2)_{\mu\mu}{}^\nu\partial_\nu)(x_0)
      |_{\varepsilon=0}.\cr}$$
The first two assertions now follow. We use
\cite{\refGib, Lemma 1.9.3} to see that the asymptotic series of the
variation is the variation of the asymptotic series. We equate coefficients
in the following two asymptotic expansions to complete the proof:
$$\eqalign{
    &\tsize\sum_n\partial_\varrho
    a_n(1,D(\varrho),\DB )t^{(n-m)/2}
    \sim\partial_\varrho\PTR(e^{-tD(\varrho)_\DB })\cr
    =&\PTR(-t\partial_\varrho D(\varrho)e^{-tD(\varrho)_\DB })\cr
    \sim&-\tsize\sum_k a_k(\partial_\varrho
     D(\varrho),D(\varrho),\DB )t^{(k+2-m)/2}.\qed\cr}$$\enddemo

To use Lemma \BREFg, we shall need some variational formulas.
Let $R_{\mu\nu\sigma}{}^\delta$ be the curvature of the
Levi-Civita connection with the sign convention that the Ricci tensor
is given by $\rho_{\nu\sigma}:=R_{\mu\nu\sigma}{}^\mu$ and the scalar
curvature is
given by
$\tau:=g^{\nu\sigma}\rho_{\nu\sigma}$. Let
$\Delta_0=\delta d$ be the scalar Laplacian, let $\dvol$ be the Riemannian
measure,
and let
$\Cur_{\mu\nu}$ be the curvature tensor of
$\nabla$.
\proclaim{\BREFh\ Lemma} Let $\nabla(\varepsilon):=\nabla+\varepsilon q_1$
and $g(\varrho):=g+\varrho q_2$. Let
$\DF_{ij}:=(\partial_\varepsilon|_{\varepsilon=0}F)_{ij}$,
$\DR_{\mu\nu\sigma}{}^\delta:
=(\partial_\varrho|_{\varrho=0}R)_{\mu\nu\sigma}{}^\delta$,
and let $\DD :=\partial_\varrho|_{\varrho=0}\Delta_0$. Then\roster
\smallskip\item $\DF_{ij}=q_{1,j;i}-q_{1,i;j}$ and
   $\partial_\varepsilon|_{\varepsilon=0}(\nabla\Cur)_{ij;k}
    =\DF_{ij;k}+[q_{1,k},\DF_{ij}]$.
\smallskip\item
$\DD f=q_{2,ij}f_{;ij}+(2q_{2,ij;j}-q_{2,jj;i})f_{;i}/2$.
\smallskip\item $\partial_\varrho\dvol|_{\varrho=0}=q_{2,ii}\dvol/2$.
\smallskip\item $(\partial_\varrho\Gamma)_{\mu\nu}{}^\sigma|_{\varepsilon=0}
     =g^{\sigma\gamma}(
    q_{2,\mu\gamma;\nu}+q_{2,\nu\gamma;\mu}-q_{2,\mu\nu;\gamma})/2$.
\smallskip\item
 $\DR_{\mu\nu\sigma}{}^\delta=g^{\delta\gamma}
    (q_{2,\mu\sigma;\gamma\nu}+q_{2,\nu\gamma;\sigma\mu}
    -q_{2,\mu\gamma;\sigma\nu}-q_{2,\nu\sigma;\gamma\mu}
    -q_{2,\sigma\beta}R_{\mu\nu\gamma}{}^\beta$
\smallskip\item " "\quad
    $-q_{2,\gamma\beta}R_{\mu\nu\sigma}{}^\beta)/2$.
\smallskip\item $\partial_\varrho|_{\varrho=0}\tau
   =-q_{2,ij}\rho_{ij}+\DR_{kiik}$.
\smallskip\item $\partial_\varrho|_{\varrho=0}|\rho|^2=
   2\DR_{kijk}\rho_{ij}-2q_{2,ij}\rho_{ik}\rho_{jk}$.
\smallskip\item $\partial_\varrho|_{\varrho=0}|R|^2=
   2\DR_{ijkl}R_{ijkl}-2q_{2,jn}R_{ijkl}R_{inkl}$.
\endroster\endproclaim

\demo{Proof} The assertion (1) is immediate from the definition.
Assertion (2)
follows from Lemma \BREFg. Assertions (3) and (4) are straightforward
calculations.
Assertion (5) follows from assertion (4) and from the identity
\footnote{We are grateful to Arkady Tseytlin who pointed out
a sign error in this identity in the previous version of the paper}:
$$\eqalign{
  &q_{2,\sigma\gamma;\nu\mu}-q_{2,\sigma\gamma;\mu\nu}=
  - q_{2,\gamma\rho}R_{\mu\nu\sigma}{}^\rho -
   q_{2,\sigma\rho}R_{\mu\nu\gamma}{}^\rho.\cr}$$
Raising and lowering indices does not
commute with varying the metric so we emphasize that the tensor $\DR$ is
the variation of a tensor of type $(3,1)$. The remaining assertions now
follow.
\qed\enddemo

\head\S3 Manifolds without boundary\endhead

Lemma \BREFg\ reduces the computation of $a_n(Q,D)$ to the special cases
$a_n(Q_i,D)$ for $i=0,1,2$. Recall that $a_n(Q,D)=0$ for $n$ odd;
$a_{-2}(Q,D)=0$
if $\ord(Q)\le1$. If
${\Cal P}$ is a scalar invariant, let ${\Cal P}[M]:=\textstyle\int_M{\Cal
P}(x)\dvol(x)$. Let $\ptr$ be the fiber trace. We refer to Gilkey
\cite{\refGia} for the proof of the following result:

\proclaim{\CREFa\ Theorem} Let $M$ be a compact Riemannian manifold
without boundary, let $D$ be an operator of Laplace type, and let
$Q_0\in\pend(V)$.
Then
\roster
\smallskip\item $a_{0}(Q_0,D)=(4\pi)^{-m/2}\ptr\{Q_0\}[M]$.
\smallskip\item $a_2(Q_0,D)
         =(4\pi)^{-m/2}6^{-1}\ptr\{Q_0(6E+\tau)\}[M]$.
\smallskip\item
      $a_4(Q_0,D)=(4\pi)^{-m/2}360^{-1}\ptr\{Q_0(60E_{;kk}
    +60\tau E+180E^2$
    \smallskip\item " "\qquad
    $+12\tau_{;kk}+5\tau^2-2 |\rho |^{ 2}
    +2 | R |^2+30\Cur_{ ij}\Cur_{ij})\}[M]$.
\smallskip\item $a_{6}(Q_0,D)=(4\pi)^{-m/2}\ptr\bigl\{Q_0/7!(
    18\tau_{;ii jj}+17\tau_{;k}\tau_{;k}
    -2\rho_{ij;k}\rho_{ ij;k}$
    \smallskip\item " "\qquad $-4\rho_{jk;n}\rho_{jn;k}
    +9R_{ij kl;n}R_{ij kl;n}+28\tau\tau_{;nn}
    -8\rho_{jk}\rho_{jk;nn}$
    \smallskip\item " "\qquad
    $+24\rho_{ jk}\rho_{jn;kn}
    +12R_{ij k\ell}R_{ij k\ell ;nn}
    +35/9\tau^{3}
    -14/3\tau\rho^2$
     \smallskip\item " "\qquad
    $+14/3\tau R^2
     -208/9\rho_{jk}\rho_{jn}\rho_{kn}
     -64/3\rho_{ ij}\rho_{kl}R_{ik jl}$
     \smallskip\item " "\qquad
     $-16/3\rho_{jk}R_{jn \ell i}R_{kn \ell i}
     -44/9R_{ij kn}R_{ij \ell p}R_{kn \ell p}$
     \smallskip\item " "\qquad
     $-80/9R_{ij kn}R_{i\ell  kp}R_{j\ell  np})
     +360^{-1}Q_0(
      8\Cur_{ij;k} \Cur_{ij;k}
     +2\Cur_{ij;j}\Cur_{ ik;k}$
     \smallskip\item " "\qquad
     $+6\Cur_{ij;kk}\Cur_{ij}
     +6\Cur_{ij}\Cur_{ij;kk}
     -12\Cur_{ij}\Cur_{jk}\Cur_{ki}
     -6R_{ij kn}\Cur_{ij}\Cur_{kn}$
     \smallskip\item " "\qquad
      $-4\rho_{jk}\Cur_{jn}\Cur_{kn}
     +5\tau\Cur_{kn}\Cur_{kn}
     +6E_{;ii jj}+30EE_{;ii}
     +30E_{;ii}E$
     \smallskip\item " "\qquad
     $+30E_{;i}E_{;i}
     +60E^{3}+12E\Cur_{ij}\Cur_{ij}
     +6\Cur_{ij}E\Cur_{ij}
     +12\Cur_{ij}\Cur_{ij}E$
     \smallskip\item " "\qquad
     $+10\tau E_{;kk}+4\rho_{jk}E_{;jk}
     +12\tau_{;k}E_{;k}
     -6E_{;j}\Cur_{ij;i}
     +6\Cur_{ij;i}E_{;j}$
     \smallskip\item " "\qquad
     $+30EE\tau +12E\tau_{;kk}+5E\tau^2
     -2E\rho_{ jk}\rho_{jk}+2ER^2)\}[M]$.
\endroster\endproclaim

Next, we study the invariants $a_n(Q_1,D)$.

\proclaim{\CREFb\ Theorem} Let $M$ be a compact Riemannian manifold
without boundary, let $D$ be an operator of Laplace type, and let
$Q_1=\partial_\varepsilon D(g,\nabla+\varepsilon q_1,E)|_{\varepsilon=0}$.
Then
\roster
\smallskip\item $a_0(Q_1,D)=0$.
\smallskip\item
   $a_2(Q_1,D)=-(4\pi)^{-m/2}360^{-1}\ptr\{60\Cur_{ij}\DF_{ij}\}[M].$
\smallskip\item
    $a_4(Q_1,D)=-(4\pi)^{-m/2}360^{-1}\ptr\{-8\Cur_{ij;k}\DF_{ij;k}
    -8\Cur_{ij;k}q_{1,k}\Cur_{ij}$
\smallskip\item " "\qquad $+8\Cur_{ij;k}\Cur_{ij}q_{1,k}
     +4\Cur_{ij;j}\DF_{ik;k}
     +4\Cur_{ij;j}q_{1,k}\Cur_{ik}
    -4\Cur_{ij;j}\Cur_{ik}q_{1,k}$
\smallskip\item " "\qquad $-36\Cur_{ij}\Cur_{jk}{}\DF_{ki}
     -12R_{ijkn}\Cur_{ij}{}\DF_{kn}
     -8\rho_{jk}\Cur_{jn}\DF_{kn}
     +10\tau\Cur_{kn}\DF_{kn}$
\smallskip\item " "\qquad
    $-60E_{;k}q_{1,k}E+60E_{;k}Eq_{1,k}+30E\Cur_{ij}\DF_{ij}
     +30E\DF_{ij}\Cur_{ij}\}[M]$.
\endroster\endproclaim

\demo{Proof} We use Lemma \BREFg\ and Theorem
\CREFa. Note that
$\Delta_0(f)=-f_{;kk}$ is independent of the connection
$\nabla$ for $f\in C^\infty(M)$. We compute:
\roster
\smallskip\item $\partial_\varepsilon\ptr(60E_{;kk})|_{\varepsilon=0}
  =60\partial_\varepsilon\ptr(E)_{;kk}|_{\varepsilon=0}
  =-60\partial_\varepsilon\Delta_0\ptr(E)=0$.
\smallskip\item
$\partial_\varepsilon\ptr(30\Cur ^2)|_{\varepsilon=0}
   =\ptr(60\Cur_{ij}\DF_{ij})$.
\smallskip\item
$\partial_\varepsilon\ptr(8\Cur_{ij;k}\Cur_{ij;k}+6\Cur_{ij;kk}
   \Cur_{ij}+6\Cur_{ij}\Cur_{ij;kk})|_{\varepsilon=0}$
   \smallskip\item " "\qquad
    $=\partial_\varepsilon\{\ptr(-4\Cur_{ij;k}\Cur_{ij;k})+
     \ptr(6\Cur_{ij}\Cur_{ij})_{;kk}\}|_{\varepsilon=0}$
   \smallskip\item " "\qquad $=\ptr(-8\Cur_{ij;k}\DF_{ij;k}
    -8\Cur_{ij;k}q_{1,k}\Cur_{ij}
    +8\Cur_{ij;k}\Cur_{ij}q_{1,k})$.
\smallskip\item $\partial_\varepsilon\ptr(30EE_{;ii}
       +30E_{;ii}E+30E_{;i}E_{;i})|_{\varepsilon=0}$
    \smallskip\item " "\qquad
     $=\partial_\varepsilon\{\ptr(-30E_{;i}E_{;i})
    +\ptr(30EE)_{;ii}\}|_{\varepsilon=0}$
    \smallskip\item " " \qquad $=\ptr(-60E_{;k}q_{1,k}E+60E_{;k}Eq_{1,k})$.
\qed
\endroster\enddemo

We conclude this section by studying $a_n(Q_2,D)$. The following result
is a consequence of Lemmas \BREFg, Lemma \BREFh, and Theorem \CREFa. We
omit the formula for $a_4(Q_2,D)$ in the interests of brevity.

\proclaim{\CREFc\ Theorem} Let $M$ be a compact Riemannian manifold
without boundary, let $D$ be an operator of Laplace type on
$C^\infty(V)$, and let
$Q_2=\partial_\varepsilon D(g+\varepsilon q_2,\nabla,E)|_{\varepsilon=0}$.
\roster
\smallskip\item $a_{-2}(Q_2,D)=-\frac 12 a_0(q_{2,ii},D)$.
\smallskip\item $a_{0}(Q_2,D)=-\frac 12 a_2(q_{2,ii},D)
     -(4\pi)^{-m/2}6^{-1}\ptr\{(-q_{2,ij}\rho_{ij}+\DR_{kiik})I_V\}[M]$.
\smallskip\item $a_2(Q_2,D)=-\frac 12 a_4(q_{2,ii},D)
       -(4\pi)^{-m/2}360^{-1}\bigl\{-\DD \ptr(60E+12\tau I_V)$
\smallskip\item " "\qquad
       $+\ptr\{10(-q_{2,ij}\rho_{ij}+\DR_{kiik})\tau I_V
        -2(2\DR_{kijk}\rho_{ij}-2q_{2,ij}\rho_{ik}\rho_{jk})I_V$
\smallskip\item " "\qquad
        $+2(2\DR_{ijkl}R_{ijkl}-2q_{2,jn}R_{ijkl}R_{inkl})I_V
      +60(-q_{2,ij}\rho_{ij}+\DR_{kiik})E$
\smallskip\item " "\qquad $-60q_{2,ik}\Cur_{ij}\Cur_{ik}\bigr\}[M]$.
\endroster\endproclaim

\head \S4 Manifolds with boundary\endhead

We now suppose $M$ has smooth non-empty boundary
$\partial M$. Near $\partial M$, let $e_i$ be a local orthonormal frame
for $TM$
where we normalize the choice so that $e_m$ is the inward unit normal.
We let indices
$a,b,...$ range from $1$ through $m-1$ and index the resulting
orthonormal frame for
the tangent bundle $T(\partial M)$ of the boundary. Let $f\in C^\infty(V)$.
Let $S$
be an endomorphism of $V$ defined on
$\partial M$. The Neumann boundary operator is defined by
$B_S^+f:=(\nabla_m+S)f|_{\partial M}$ and the Dirichlet boundary
operator is defined by $B_S^-f=f|_{\partial M}$; we set
$S=0$ with Dirichlet boundary conditions to have a uniform notation. Let
$L_{ab}=(\nabla_{e_a}e_b,e_m)$ be the second fundamental form. Let `:' denote
multiple covariant differentiation tangentially with respect to the
Levi-Civita connection of the metric on the boundary and
the connection $\nabla$ on $V$. The difference between `;' and `:' is given by
the second fundamental form. For example,
$E_{;a}$ and
$E_{:a}$ agree since there are no tangential indices in
$E$ to be differentiated while $E_{;ab}=E_{:ab}-L_{ab}E_{;m}$.
There are some new features here which are not
present in the case of manifolds without boundary in the following
formulas.  The invariants $a_n(Q,D,\DB )$ are non-zero for odd $n$ and
the normal derivatives of $Q_0$ enter. We still have
$a_{-2}(Q,D,\DB )=0$ if $\ord(Q)\le1$.

\proclaim{\DREFa\ Theorem} Let $M$ be a compact Riemannian
manifold with smooth boundary, let $D$ be an operator of
Laplace type and let $Q_0\in\pend(V)$. Then\roster
\smallskip\item $a_0(Q_0,D,\DB_S^\pm)
   =(4\pi)^{-m/2}\ptr\{Q_0\}[M].$
\smallskip\item $a_1(Q_0,D,\DB_S^\pm)
   =\pm(4\pi)^{-(m-1)/2}4^{-1}\ptr\{Q_0\}[\partial M].$
\smallskip\item $a_2(Q_0,D,\DB_S^\pm)
   =(4\pi)^{-m/2}6^{-1}\bigl\{\ptr\{Q_0(6E+\tau)\}[M]$
\smallskip\item " "\qquad $+\ptr\{Q_0(2L_{aa}+12S)
    +(3^+,-3^-)Q_{0;m}\}[\partial M]\bigr\}.$
\smallskip\item $a_3(Q_0,D,\DB_S^\pm)
    =(4\pi)^{-(m-1)/2}{384}^{-1}
    \ptr\{Q_0((96^+,-96^-)E+(16^+,-16^-)\tau$
    \smallskip\item " "\qquad $+(8^+,-8^-)R_{amam}
    +(13^+,-7^-)L_{aa}L_{bb}+(2^+,10^-)L_{ab}L_{ab}
    +96SL_{aa}$
    \smallskip\item " "\qquad $+192S^{2})
    +Q_{0;m}((6^+,30^-)L_{aa}+96S)
    +(24^+,-24^-)Q_{0;mm}\}[\partial M]$.
\smallskip\item $a_4(Q_0,D,\DB_S^\pm)
    =(4\pi)^{-m/2}360^{-1}\big\{
    \ptr\{Q_0(60E_{;kk}+60\tau
E+180E^{2}+30\Cur_{ij}\Cur_{ij}$
   \smallskip\item " "\qquad $+12\tau_{;kk}+5\tau^{2}-2\rho^{2}
    +2R^{2})\}[M]$
    \smallskip\item " "\qquad$+\ptr\{Q_0(240^+,-120^-)E_{;m}
     +(42^+,-18^-)\tau_{;m}+24L_{aa:bb}
     +120EL_{aa}$
    \smallskip\item " "\qquad$+20\tau L_{aa}+4R_{am am}L_{bb}
     -12R_{ambm}L_{ab}+4R_{abcb}L_{ac}+360(SE+ES)$
    \smallskip\item " "\qquad$
     +21^{-1}\{(280^+,40^-)L_{aa}L_{bb}L_{cc}
     +(168^+,-264^-)L_{ab}L_{ab}L_{cc}$
    \smallskip\item " "\qquad$
     +(224^+,320^-)L_{ab}L_{bc}L_{ac}\}
     +120S\tau+144SL_{aa}L_{bb}
     +48SL_{ab}L_{ab}$
    \smallskip\item " "\qquad$+480S^{2}L_{aa}+480S^{3}
     +120S_{:aa}
     +Q_{0;m}((180^+,-180^-)E+(30^+,-30^-)\tau$
    \smallskip\item " "\qquad$
     +(84^+,-180^-)/7\cdot L_{aa}L_{bb}
     +(84^+,60^-)/7\cdot L_{ab}L_{ab}
     +72SL_{aa}+240S^{2}$
    \smallskip\item " "\qquad$
     +Q_{0;mm}(24L_{aa}+120S)+(30^+,-30^-)Q_{0;iim}
     \}[\partial M]\big\}.$
\endroster\endproclaim

\proclaim{\DREFb\ Theorem} Let $M$ be a compact Riemannian
manifold with smooth boundary, let $D$ be an operator of
Laplace type, and let
$Q_1=\partial_{\varepsilon} D(g,\nabla+\varepsilon q_1,E)|_{\varepsilon=0}$.
Then
\roster
\smallskip\item $a_{-1}(Q_1,D,\DB_S^\pm)=0$.
\smallskip\item $a_0(Q_1,D,\DB_S^\pm)=
 -(4\pi)^{-m/2}6^{-1} \ptr ((-12^+,0^-)q_{1,m})[\partial M]$.
\smallskip\item $a_1(Q_1,D,\DB_S^\pm)=-(384)^{-1} (4\pi)^{-(m-1)/2}
     \ptr ((-96^+,0^-)q_{1,m}L_{aa}$
\smallskip\item " "\qquad $-384Sq_{1,m})[\partial M]$.
\smallskip\item $a_2(Q_1,D,\DB_S^\pm)=-(4\pi)^{-m/2}360^{-1}\{
    \ptr(60\Cur_{ij}\DF_{ij})[M]$
\smallskip\item " "\qquad$+\ptr\{(-720^+,0^-)q_{1,m}E
     +(-120^+,0^-)q_{1,m}\tau+(-144^+,0^-)q_{1,m}L_{aa}L_{bb}$
\smallskip\item " "\qquad$+(-48^+,0^-)q_{1,m}L_{ab}L_{ab}
      -960Sq_{1,m}L_{aa}
      -1440S^2q_{1,m}) [\partial M] \}$.
\smallskip\item
     $a_3(Q_1,D,\DB ^-)=5760^{-1}(4\pi)^{(m-1)/2}\ptr\{
       240\Cur_{ab}\DF_{ab}-720\Cur_{am}\DF_{am}\}[\partial M]$.
\smallskip\item If the boundary of $M$ is totally geodesic, then
\smallskip\item " "
     $a_3(Q_1,D,\DB_S^\pm)=-5760^{-1}(4\pi)^{(m-1)/2}
     \ptr(-1440E_{;m}q_{1,m}$
\smallskip\item " "\qquad $+1440(q_{1,m}E-Eq_{1,m})S
       +240\Cur_{ab}\DF_{ab}-960\tau Sq_{1,m}-240\rho_{mm}Sq_{1,m}$
\smallskip\item " "\qquad $ +180\Cur_{am}\DF_{am}
       -270\tau_{;m}q_{1,m}+720S_{:a}q_{1,m:a}$
\smallskip\item " "\qquad $ +720S_{:a}(q_{1,a}S-Sq_{1,a})
      -2880E(Sq_{1,m}+q_{1,m}S) -5760q_{1,m}S^3)[\partial M]$.
\endroster\endproclaim

\demo{Proof} If $Q_0=q_0I_V$ for $q_0\in C^\infty(M)$ is a scalar operator,
then
Theorem
\DREFa\ follows from Branson and Gilkey \cite{\refBGa}. If $Q_0$ is not a
scalar
operator, we must worry about the lack of commutativity; the only point
at which this
enters is in the coefficient of $\ptr(Q_0SE)$ and $\ptr(Q_0ES)$. We express
$$a_4(Q_0,D,\DB_S)=(4\pi)^{-m/2}360^{-1}\ptr(C_1Q_0SE+C_2Q_0ES)[\partial M]+
\text{other terms};$$
the sum
$C_1+C_2=720$ is determined by the scalar case. If  $D$, $Q_0$, and $S$
are real,
then $\PTR(Q_0e^{-tD})$ is real; this shows that $C_1$ and $C_2$ are real.
If $Q_0$
$D$, and $S$ are self-adjoint, $\PTR(Q_0e^{-tD})$ is real
so $\ptr(Q_0(C_1SE+C_2ES))[\partial M]$ is real; this now shows $C_1=C_2$
and completes the proof of Theorem \DREFa.

To keep boundary conditions constant, let
$S(\varepsilon):=S-\varepsilon q_{1,m}$ so
$\partial_\varepsilon S|_{\varepsilon=0}=-q_{1,m}$. Assertions (1)-(5)
of Theorem
\DREFb\ now follow directly from Lemma \BREFg, from Lemma \BREFh, and
from Theorem
\DREFa. In
\cite{\refBGV}, we showed that
\roster\smallskip\item " "
   $a_{5}(1,D,\DB_S^\pm)
   =\pm 5760^{-1}(4\pi)^{
   (m-1)/2}\ptr\{(360E_{;mm}+1440E_{;m}S$
\item " "\qquad
   $+720E^2+240E_{:aa}+240\tau E+120\Cur_{ab}\Cur_{ab}
   +48\tau_{;ii}+20\tau^{ 2}-8\rho^{2}$
\item " "\qquad $+8R^{2 }
   -120\rho_{mm}E-20\rho_{mm}\tau +480\tau S^{2}
    +(90^{+},-360^{-})\Cur_{ am}\Cur_{am}$
\item " "\qquad $+12\tau_{;mm}
   +24\rho_{mm:aa}+15\rho_{mm;mm}
   +270\tau_{;m}S+120\rho_{mm}S^{2}+960S_{:aa}S$
\item " "\qquad
   $+600S_{:a}S_{:a}+16R_{am mb}\rho_{ab}
    -17\rho_{mm}\rho_{mm}-10R_{ am mb}R_{am mb }
    +2880ES^{2}$
\item " "\qquad
   $+1440S^{4})+\DE \} [\partial M]$\endroster
The variation of the terms other than $\DE $ gives rise to the expressions
listed
in Theorem \DREFb\ (5,6). The remainder term
$\DE $ is given below. It vanishes if the boundary is totally geodesic and
involves 40 undetermined coefficients.
\roster\item" "
   $\DE =d_1^\pm L_{aa}E_{;m}+d_2^\pm L_{aa}\tau_{:m}
   +d_3^\pm L_{ab}R_{ammb;m}+d_4^+L_{aa}S_{:bb}
   +d_5^+L_{ab}S_{:ab}$
\item " "\qquad $+d_6^+L_{aa:b}S_{:b}
   +d_7^+L_{ab:a}S_{:b}+d_8^+L_{aa:bb}S+d_9^+L_{ab:ab}S
   +d_{10}^\pm L_{aa:b}L_{cc:b}$
\item " "\qquad
   $+d_{11}^\pm L_{ab:a}L_{cc:b}
    +d_{12}^\pm L_{ab:a}L_{bc:c}
    +d_{13}^\pm L_{ab:c}L_{ab:c}
    +d_{14}^\pm L_{ab:c}L_{ac:b}$
\item " "\qquad
    $+d_{15}^\pm L_{aa:bb}L_{cc}
    +d_{16}^\pm L_{ab:ab}L_{cc}
    +d_{17}^\pm L_{ab:ac}L_{bc}
    +d_{18}^\pm L_{aa:bc}L_{bc}
    +d_{19}^\pm L_{bc:aa}L_{bc}$
\item " "\qquad
    $+1440^+L_{aa}SE
    +d_{20}^+L_{aa}S\rho_{mm}
    +240^+ L_{aa}S\tau
    +d_{21}^+L_{ab}S\rho_{ab}$
\item " "\qquad
    $+d_{22}^+L_{ab}SR_{mabm}
    +(195^+,105^-)L_{aa}L_{bb}E
    +(30^+,150^-)L_{ab}L_{ab}E$
\item " "\qquad
    $+(195^+/6,105^-/6)L_{aa}L_{bb}\tau
    +(5^+,25^-)L_{ab}L_{ab}\tau
    +d_{23}^\pm L_{aa}L_{bb}\rho_{mm}$
\item " "\qquad
    $+d_{24}^\pm L_{ab}L_{ab}\rho_{mm}
    +d_{25}^\pm L_{aa}L_{bc}\rho_{bc}
    +d_{26}^\pm L_{aa}L_{bc}R_{mbcm}
    +d_{27}^\pm L_{ab}L_{ac}\rho_{bc}$
\item " "\qquad
    $+d_{28}^\pm L_{ab}L_{ac}R_{mbcm}
    +d_{29}^\pm L_{ab}L_{cd}R_{acbd}
    +d_{30}^+ L_{aa}S^{3}
    +d_{31}^+ L_{aa}L_{bb}S^{2}$
\item " "\qquad
    $+d_{32}^+ L_{ab}L_{ab}S^{2}
    +d_{33}^+ L_{aa}L_{bb}L_{cc}S
    +d_{34}^+ L_{aa}L_{bc}L_{bc}S
    +d_{35}^+ L_{ab}L_{bc}L_{ac}S$
\item " "\qquad
    $+d_{36}^\pm L_{aa}L_{bb}L_{cc}L_{dd}
    +d_{37}^\pm L_{aa}L_{bb}L_{cd}L_{cd}
    +d_{38}^\pm L_{ab}L_{ab}L_{cd}L_{cd}$
\item " "\qquad
    $+d_{39}^\pm L_{aa}L_{bc}L_{cd}L_{db}
    +d_{40}^\pm L_{ab}L_{bc}L_{cd}L_{da}$
\endroster
The variation of $\DE$ is zero for Dirichlet boundary conditions or if the
boundary is totally geodesic. \qed\enddemo

To study $a_n(Q_2,D,\DB_S^\pm)$ we need some additional formulas.

\proclaim{\DREFc\ Lemma}
\roster\smallskip\item Let $g(\varrho):=g+\varrho q_2$,
$\DN:=\partial_\varrho|_{\varrho=0}e_m(\varrho)$, and
$\DL_{\alpha\beta}:=\partial_\varrho|_{\varrho=0}L_{\alpha\beta}$. Then
\smallskip\item " "\quad{\rm a)} $\DN=-q_{2,am}e_a-q_{2,mm}e_m/2$.
\smallskip\item " "\quad{\rm b)} $\DL_{ab}=(q_{2,am;b}
     +q_{2,bm;a}-q_{2,ab;m}-q_{2,mm}L_{ab})/2$.
\smallskip\item $q_{2,am;a}=q_{2,am:a}-L_{aa}q_{2,mm}+L_{ab}q_{2,ab}$.
\smallskip\item $\DR_{kiik}=q_{2,ki;ki}-q_{2,ii;kk}$.
\endroster\endproclaim

\demo{Proof} Let $1\le\alpha,\beta\le m-1$. Let $y=(y^\alpha)$ be local
coordinates on
the boundary $\partial M$ centered at $y_0$. We suppose
$g_{\alpha\beta}(y_0)=\delta_{\alpha\beta}+O(|y|^2)$. Introduce
coordinates
$x=(y,x^m)$ so the curves $t\mapsto(y,t)$ are unit speed geodesics
perpendicular
to the boundary. Then
$g_{mm}=1$ and $g_{\alpha m}=0$; $\partial_m$ is the inward geodesic normal
vector field for $g$. Let $N(\varrho)$ be the inward geodesic normal
vector field
for the metric $g(\varrho)$. Expand
$N(\varrho)(y_0)=\partial_m+\varrho(c^m\partial_m
  +c^\beta\partial_\beta)+O(\varrho^2)$. We prove the first assertion
by solving
the equations
$$\eqalignno{
  0=&g(\varrho)(N(\varrho),\partial_\alpha)(y_0)
   =\varrho(c^\alpha+q_{2,\alpha m})(y_0)+O(\varrho^2)\cr
  1=&g(\varrho)(N(\varrho),N(\varrho))(y_0)
   =1+\varrho(2c^m+q_{2,mm})(y_0)+O(\varrho^2)\cr}$$
to see $c^\alpha(y_0)=-q_{2,m\alpha}(y_0)$ and $c^m(y_0)=-q_{2,mm}/2$.
We use Lemma
\BREFh\ to compute the variation of the Christoffel symbols and complete
the proof by
computing:
$$\eqalign{
L_{\alpha\beta}=&\Gamma(\rho)_{\alpha\beta}{}^ig(N(\varrho),\partial_i)\cr
\DL_{\alpha\beta}(y_0)=&(\dot\Gamma_{\alpha\beta}{}^m
   -q_{2,mm}L_{\alpha\beta}/2)(y_0).\cr}$$
The second assertion is immediate, the third follows from Lemma \BREFh.
\qed \enddemo

Dirichlet boundary conditions are unchanged by a variation of the metric
$g$. The
following result follows from Lemma \BREFg, Theorem \DREFa, and Lemma
\DREFc. We omit the formula for $a_2$ in the interests of brevity.

\proclaim{ \DREFd\ Theorem} Let $M$ be a compact Riemannian manifold with
smooth
boundary, let $D$ be an operator of Laplace type, and let
$Q_2:=\partial_\varepsilon|_{\varepsilon=0}D(g+\varepsilon q_2,\nabla,E)$.
Then
\roster
\smallskip\item $a_{-2}(Q_2,D,\DB^-)
   =-(4\pi)^{-m/2}\ptr\{q_{2,ii}/2\}[M]$.
\smallskip\item $a_{-1}(Q_2,D,\DB^-)
   =(4\pi)^{-(m-1)/2}4^{-1}\ptr\{q_{2,aa}/2\}[\partial M]$.
\smallskip\item $a_{0}(Q_2,D,\DB^-)
    =-(4\pi)^{-m/2}6^{-1}\bigl\{\ptr\{q_{2,ii}(6E+\tau)/2-q_{2,ij}\rho_{ij}$
\smallskip\item " "\qquad $+\DR_{kiik}\}[M]$
      $+\ptr\{q_{2,aa}L_{aa}+2\DL_{aa}-2q_{2,ab}L_{ab}\}[\partial M]
      \bigl\}$.
\smallskip\item $a_{1}(Q_2,D,\DB^-)=(4\pi)^{-(m-1)/2}{384}^{-1}\ptr\{
      q_{2,aa}(96E+16\tau+8R_{amam}$
\smallskip\item " "\qquad $+7L_{aa}L_{bb}-10L_{ab}L_{ab})/2
      +16(-q_{2,ij}\rho_{ij}+\DR_{kiik})$
\smallskip\item " "\qquad $+8(\DR_{amam}-q_{2,ab}R_{ambm}+2R_{ama\DN})$
                          $+7(2\DL_{aa}L_{bb}-2q_{2,ab}L_{ab}L_{cc})$
\smallskip\item " "\qquad $-10(2\DL_{ab}L_{ab}-2q_{2,ac}L_{ab}L_{cb})
                   \}[\partial M]$.
\smallskip\item Suppose for simplicity that the metric $g$ is flat, i.e. that
$R_{ijkl}=0$. Then
\smallskip\item " "$a_2(Q_2,D,{\Cal B}^-)=
        -(4\pi)^{-m/2}360^{-1}\{
        -{\Cal D}\ptr(60E)+\ptr (60\DR_{ijji}E$
\smallskip\item " "\qquad
     $+30q_{2,ii}E_{;kk}+90q_{2,ii}E^2+15q_{2,ii}\Omega^2)\}[M]$
  \smallskip\item " "\qquad
$-(4\pi)^{-m/2}360^{-1}\ptr\{60E_{;m}q_{2,mm} +120E_{;a}q_{2,am}
       -18\DR_{ikki;m}$
  \smallskip\item " "\qquad $+L_{aa}(20\DR_{ikki}
         +4\DR_{bmbm})+12\DR_{ambm}L_{ab}+4\DR_{abcb}L_{ac} $
  \smallskip\item " "\qquad $+q_{2,dd}(-60E_{;m}+60EL_{aa}
         +12L_{aa:bb} $
  \smallskip\item " "\qquad $+20/21L_{aa}L_{bb}L_{cc}
         +44/7L_{ab}L_{ab}L_{cc}+160/21L_{ab}L_{bc}L_{ac})$
  \smallskip\item " "\qquad $(120E+40/7L_{aa}L_{bb}+88/7L_{ab}L_{ab})
           (\DL_{cc}-q_{2,cd}L_{cd})$
  \smallskip\item " "\qquad $+88/7(
           2\DL_{ab}L_{ab}L_{cc}-2q_{2,bd}L_{ab}L_{ad}L_{cc})$
  \smallskip\item " "\qquad $+320/7(\DL_{ab}L_{bc}L_{ca}
           -q_{2,ad}L_{ab}L_{bc}L_{cd}) +12 L_{bb:c}q_{2,aa:c}\}
            [\partial M]$.
\endroster\endproclaim

We now study Neumann boundary conditions. The situation is quite different as
Neumann boundary conditions are {\bf not} invariant under general
perturbations of
the metric; if $q_{am}\ne0$ on $\partial M$, $\DB^+_S(\varrho)$ will involve
tangential derivatives regardless of how $S$ is varied.
Thus Lemma \BREFg\ is not
directly applicable. Nevertheless, we can still compute the first three
terms in the
asymptotic expansion.

\proclaim{\DREFe\ Theorem} Let $M$ be a compact Riemannian manifold with
smooth
boundary, let $D$ be an operator of Laplace type, and let
$Q_2:=\partial_\varepsilon D(g+\varepsilon q_2,\nabla,E)|_{\varepsilon=0}$.
Then
\roster\smallskip\item $a_{-2}(Q_2,D,\DB^+_S)
   =-(4\pi)^{-m/2}\ptr\{q_{2,ii}/2\}[M]$.
\smallskip\item $a_{-1}(Q_2,D,\DB^+_S)
   =-(4\pi)^{-(m-1)/2}4^{-1}\ptr\{q_{2,aa}/2\}[\partial M]$.
\smallskip\item $a_{0}(Q_2,D,\DB^+_S)
    =-(4\pi)^{-m/2}6^{-1}\ptr\{q_{2,ii}(6E+\tau)/2
     -q_{2,ij}\rho_{ij}\}[M]$
\smallskip\item " "\qquad
    $-(4\pi)^{-m/2}6^{-1}\ptr\{q_{2,aa}L_{bb}-q_{2,ab}L_{ab}
        -6q_{2,mm}S+6q_{2,aa}\}[\partial M]$.
\endroster\endproclaim
\demo{Proof} Define $\ord(q_{2,ij})=0$, $\ord(E)=2$, $\ord(R_{ijkl})=2$,
$\ord(\Cur)=2$, $\ord(L)=1$, and $\ord(S)=1$. Increase the order by $1$ for
each explicit covariant derivative which is present. Dimensional analysis then
shows the interior integrands in the formula for $a_n(Q_2,D,\DB)$ are
homogeneous of order
$n+2$ while the boundary integrands are homogeneous of degree $n+1$. We use H.
Weyl's theorem to write a spanning set for the set of invariants and express:
$$\leqalignno{a_{-2}&(Q_2,D,\DB^+_S)
   =-(4\pi)^{-m/2}\ptr(b_1q_{2,ii}/2)[M]&(\DREFg)\cr
   a_{-1}&(Q_2,D,\DB^+_S)
   =-(4\pi)^{-(m-1)/2}4^{-1}\ptr(c_1q_{2,aa}/2
         +c_2q_{2,mm})[\partial M].&(\DREFh)\cr
    a_{0}&(Q_2,D,\DB^+_S)
    =-(4\pi)^{-m/2}6^{-1}\{\ptr(q_{2,ii}(6b_2E+b_3\tau)/2&(\DREFi)\cr
     &-b_4q_{2,ij}\rho_{ij})[M]
     +\ptr(c_3q_{2,aa}L_{bb}+c_4q_{2,ab}L_{ab}+c_5q_{2,mm}L_{aa}\cr
        &+c_6q_{2,mm}S+c_7q_{2,aa}S
    +c_8q_{2,mm;m}+c_9q_{2,aa;m})[\partial M]\}.\cr}$$
Product formulas then show the constants are independent of the dimension
$m$; these invariants form a basis for the integral invariants and are
uniquely determined for $m$ large. A word of explanation for the formula in
equation (\DREFi) is in order. We can integrate by parts to replace the
interior integrals $q_{2,ij;ij}$ and $q_{2,ii;jj}$ by boundary integrals of
$q_{2,mm;m}$, $q_{2,am;a}$, and $q_{2,aa;m}$. Since $q_{2,am:a}[\partial
M]=0$, we use Lemma \DREFc\ to omit the variable $q_{2,am;a}$.
If we take $\partial M$ empty, the boundary condition plays no role and
$\DR_{kiik}[M]=0$ (see equation (\DREFk) below). We use Theorem \CREFc\ to
see
$b_1=b_2=b_3=b_4=1$; this completes the proof of assertion (1) and the first
part of
assertion (3).

The $q_{2,am;*}$ variables do not appear in equations (\DREFg), (\DREFh),
and (\DREFi). Thus we may take a variation with
$q_{2,am}=0$ on $\partial M$. This is an essential simplification since it
means $N(\varrho)=g^{mm}(\varrho)^{1/2}\partial_m$. Thus the boundary
conditions do not involve any tangential derivatives. We set
$S(\varrho)=g^{mm}(\varrho)^{1/2}S$. Then the boundary condition is
preserved;
$\nabla_{N(\varrho)}+S(\varrho)=g^{mm}(\varrho)^{1/2}(\nabla_m+S)$.
We have $\partial_\varrho S(\varrho)|_{\varrho=0}=-q_{2,mm}S/2$. By Lemma
\BREFg,
$a_{-1}(Q_2,D,B_S^+)=-(4\pi)^{-(m-1)/2}4^{-1}\ptr(q_{2,aa}/2)[\partial M]$.
This shows $c_1=1$ and $c_2=0$ and completes the proof of assertion (2).

We use Lemma \BREFg\ and Theorem \DREFa\ to see:
$$\eqalign{
  &a_{0}(Q_2,D,B_S^+)=\cr
  &-(4\pi)^{-(m-1)/2}6^{-1}\bigl\{\ptr(q_{2,ii}(6E+\tau)/2-q_{2,ij}\rho_{ij}
   +\DR_{kiik})[M]\cr
   &+\ptr(q_{2,aa}(L_{bb}+6S)-2q_{2,ab}L_{ab}
    +2{\Cal L}_{aa}-6q_{2,mm}S)[\partial M]\bigr\}\cr}\tag{\DREFj}$$
We have that
$$\eqalign{
&{\DR}_{kiik}[M]=(q_{2,ki;ki}-q_{2,ii;kk})[M]
        =(-q_{2,am;a}+q_{2,aa;m})[\partial M]\cr
&\quad=(L_{aa}q_{2,mm}-L_{ab}q_{2,ab}+q_{2,aa;m})[\partial M]\cr
&2\DL_{aa}[\partial M]
  =(2q_{2,am;a}-q_{2,aa;m}-q_{2,mm}L_{aa})[\partial M]\cr
&\quad=(-3L_{aa}q_{2,mm}+2L_{ab}q_{2,ab}-q_{2,aa;m})[\partial M].\cr}
  \tag{\DREFk}$$
We use equation (\DREFk) to compare equations (\DREFi) and (\DREFj). This
shows
$$c_3=1,\ c_4=-1,\ c_5=-2,\ c_6=-6,\ c_7=6,\ c_8=0,\ c_9=0.\qed$$\enddemo


\head\S5 Operators of Dirac Type\endhead

In this section, we study the invariants $a_n(HA,A^2)$, where $M$ is a closed
manifold,
$A$ is an operator of Dirac type, and $H$ is a smooth endomorphism; we
refer to
Branson and Gilkey \cite{\refBGc} for a discussion of the case of
manifolds with
boundary. We begin with a technical result:

\proclaim{\EREFa\ Lemma}  Let $A=\gamma^\nu\partial_\nu-\psi$ be an operator
of Dirac type and let $D=A^2$ be the associated operator of
Laplace type. Let $D=D(g,\nabla,E)$ and let $H\in C^\infty\pend(V)$.
\roster
\smallskip\item $\omega_\mu=
    g_{\nu\mu}(-\gamma^\sigma\partial_\sigma\gamma^\nu
   +\psi\gamma^\nu+\gamma^\nu\psi
   +g^{\sigma\rho}\Gamma_{\sigma\rho}{}^\nu)/2$.
\smallskip\item Let $\phi:=\psi+\gamma^\nu\omega_\nu$. Then
    $A=\gamma^\nu\nabla_\nu-\phi$, and $\phi$ is invariantly defined.
\smallskip\item $\gamma_{i;j}+\gamma_{j;i}=0$.
\smallskip\item
   $E=-\psi^2+\gamma^\mu\partial_\mu\psi
     -g^{\nu \mu}(\partial_\mu\omega_\nu+\omega_\nu\omega_\mu
     -\omega_\sigma\Gamma_{\nu\mu}{}^\sigma)
     =-\gamma_i\gamma_j\Cur_{ij}/2+\gamma_i\phi_{;i}-\phi^2$.
\smallskip\item Let $q_{1,i}:=-H\gamma_i/2$, and
   $Q_0:=-H\phi-H_{;i}\gamma_i/2$, then $HA=Q_1+Q_0$.
\endroster\endproclaim

\demo{Proof} We compute
$$\eqalign{
 D=&A^2=(\gamma^\nu\partial_\nu-\psi)(\gamma^\mu\partial_\mu-\psi)\cr
  =&\gamma^\nu\gamma^\mu\partial_\nu\partial_\mu+
    (\gamma^\nu\partial_\nu\gamma^\mu
   -\gamma^\mu\psi-\psi\gamma^\mu)\partial_\mu+\psi^2-
    \gamma^\mu\partial_\mu\psi,\cr
 a^\mu=&-\gamma^\nu\partial_\nu\gamma^\mu+\gamma^\mu\psi+\psi\gamma^\mu,
   \text{ and }b=-\psi^2+\gamma^\mu\partial_\mu\psi.\cr}$$
Assertion (1) and the first assertion of (4) follows from Lemma \BREFd.
We prove
assertion (2) by computing:
$\gamma^i\nabla_i-\phi
  =\gamma^i\partial_i+\gamma^i\omega_i-\phi
  =\gamma^i\partial_i-\psi$.
We choose a system of coordinates
and a local frame so that $\Gamma(x_0)=0$ and so that $\omega(x_0)$=0.
Then at $x_0$, we have:
$$\eqalign{
   D&=(\gamma^\nu\nabla_\nu-\phi)(\gamma^\mu\nabla_\mu-\phi)\cr
       &=(\gamma^\nu\gamma^\mu+\gamma^\mu\gamma^\nu)/2\nabla_\nu\nabla_\mu
  +(\gamma^\mu\gamma^\nu{}_{;\mu}-\phi\gamma^\nu-\gamma^\nu\phi)\nabla_\nu\cr
  &\qquad-\gamma^\nu\phi_{;\nu}+\gamma^\nu\gamma^\mu\Omega_{\nu\mu}/2+
   \phi^2\cr
  &=-g^{\nu\mu}\nabla_\nu\nabla_\mu-E.\cr}$$
We equate coefficients to derive the second part of assertion (4). Choose a
coordinate system centered at $x_0$ so $g_{\mu\nu}=\delta_{\nu\mu}+O(|x|^2)$.
We use
\cite{\refBGb, Lemma 1.2} to see that we can choose a local frame for $V$ so
$\partial_\mu\gamma^\nu(x_0)=0$. Then we have that
$\omega_\mu(x_0)=(\psi\gamma_\mu+\gamma_\mu\psi)(x_0)/2$. We prove assertion
(3) by
computing at $x_0$:
$$\eqalign{
    2(\gamma_{\nu;\mu}&+\gamma_{\mu;\nu})=
     [\psi\gamma_\mu+\gamma_\mu\psi,\gamma_\nu]+[\psi\gamma_\nu+\gamma_\nu
\psi,\gamma_\mu]\cr
    =&\psi\gamma_\mu\gamma_\nu+\gamma_\mu\psi\gamma_\nu
    -\gamma_\nu\psi\gamma_\mu-\gamma_\nu\gamma_\mu\psi\cr
    &+\psi\gamma_\nu\gamma_\mu+\gamma_\nu\psi\gamma_\mu
    -\gamma_\mu\psi\gamma_\nu-\gamma_\mu\gamma_\nu\psi\cr
    =&\psi\gamma_\mu\gamma_\nu+\psi\gamma_\nu\gamma_\mu
      -\gamma_\mu\gamma_\nu\psi-\gamma_\mu\gamma_\nu\psi\cr
    =&\psi\delta_{\mu\nu}-\delta_{\mu\nu}\psi=0.\cr}$$
If we set $q_{1,i}=-H\gamma_i/2$, then
$Q_1=H\gamma_i\nabla_i+H_{;i}\gamma_i/2$
by Lemma \BREFg\ since $\gamma_{i;i}=0$.
We must therefore take $Q_0=-H\phi-H_{;i}\gamma_i/2$.
\qed\enddemo

The following theorem now follows from Theorem \CREFa, Theorem \CREFb,
and from
Lemma \EREFa.

\proclaim{\EREFb\ Theorem} Let $A=\gamma_\nu\partial_\nu-\psi$ be an
operator of Dirac
type on $C^\infty(V)$ over a closed manifold $M$. Adopt the notation
of Lemma \EREFa.
\roster
\smallskip\item $a_0(HA,A^2)=(4\pi)^{-m/2}\ptr\{Q_0\}[M]$.
\smallskip\item $a_2(HA,A^2)=(4\pi)^{-m/2}6^{-1}\ptr\{
     Q_0(\tau+6E)-\Cur_{ij}\DF_{ij}\}[M]$.
\smallskip\item
      $a_4(HA,A^2)=(4\pi)^{-m/2}360^{-1}\ptr\{Q_0(60E_{;kk}
    +60\tau E+180E^2+12\tau_{;kk}$
    \smallskip\item " "\qquad
    $+5\tau^2-2 |\rho |^{ 2}
    +2 | R |^2+30\Cur_{ ij}\Cur_{ij})
    +8\Cur_{ij;k}\DF_{ij;k}
    +8\Cur_{ij;k}q_{1,k}\Cur_{ij}$
\smallskip\item " "\qquad $-8\Cur_{ij;k}\Cur_{ij}q_{1,k}
     -4\Cur_{ij;j}\DF_{ik;k}
     -4\Cur_{ij;j}q_{1,k}\Cur_{ik}
    +4\Cur_{ij;j}\Cur_{ik}q_{1,k}$
\smallskip\item " "\qquad $+36\Cur_{ij}\Cur_{jk}{}\DF_{ki}
     +12R_{ijkn}\Cur_{ij}{}\DF_{kn}
     +8\rho_{jk}\Cur_{jn}\DF_{kn}
     -10\tau\Cur_{kn}\DF_{kn}$
\smallskip\item " "\qquad
    $+60E_{;k}q_{1,k}E-60E_{;k}Eq_{1,k}-30E\Cur_{ij}\DF_{ij}
     -30E\DF_{ij}\Cur_{ij}\}[M]$.
\endroster\endproclaim

\subhead\EREFe\ Remark\endsubhead The first two authors \cite{\refBGb,
Theorem 2.7}
computed
$a_0$ and
$a_2$ for $f$ scalar and showed that
$$\leqalignno{
a_0(fA,A^2)=&-(4\pi)^{-m/2}\ptr(f\phi)[M],\text{ and}&(\EREFg)\cr
a_2(fA,A^2)=&-(4\pi)^{-m/2}6^{-1}\ptr\{f(\phi\tau+6\phi E-
   \Cur_{ij;j}\gamma_i)\}[M].&(\EREFk)\cr}$$
For scalar $f$,
$\PTR(f_{;i}\gamma_ie^{-tA^2})=\PTR(Afe^{-tA^2}-fAe^{-tA^2})=0$ so
$a_n(f_{;i}\gamma_i,A^2)=0$ for all $n$ and
 we may replace
$Q_0$ by $-f\phi$ in performing our computations. It then follows that
Theorem
\EREFb\ (1) agrees with equation (\EREFg). We see Theorem \EREFb\ (2)
agrees with
equation (\EREFk) by using Lemma \BREFh\ and integrating by parts:
$$\eqalign{
 -\ptr&(\Cur_{ij}\DF_{ij})[M]
  =\ptr((f\gamma_j)_{;i}-(f\gamma_i)_{;j})\Cur_{ij}[M]/2\cr
  =&-\ptr(f\gamma_j\Cur_{ij;i}[M])=\ptr(f\gamma_i\Cur_{ij;j})[M].\cr}$$

\enddocument